\documentclass[pra,twocolumn]{revtex4}
\usepackage{graphicx}
\usepackage{amssymb,amsfonts,amsmath}

\def \beq {\begin{equation}}
\def \eeq {\end{equation}}
\def \tr {\rm Tr}

\begin{document}
\title{Quantum Theory of Radical-Ion-Pair Recombination: A New Physical Paradigm for Low-Magnetic-Field Effects}
\author{Iannis K. Kominis}
\email{ikominis@iesl.forth.gr}
\affiliation{Department of Physics, University of Crete, Heraklion
71103, Greece}
\affiliation{Institute of Electronic Structure and
Laser, Foundation for Research and Technology, Heraklion 71110,
Greece}

\begin{abstract}
A new paradigm emerging in the description of magnetic-sensitive
radical-ion-pair recombination reactions is presented. This
paradigm is founded on the realization that the recombination
process of radical-ion pairs is a continuous quantum measurement.
In the regime of low magnetic fields we describe the appearance of
purely quantum phenomena, that fade away as the magnetic field
increases. We will analyze the magnetic sensitivity of these
reactions under this perspective and bridge the new full quantum
theory with the existing classical reaction theory applicable at
high magnetic fields. Based on the quantum theory of recombination
we will then explain experimental observations incompatible with
classical reaction theory, in particular the effect of deuteration
on the magnetic sensitivity of radical-ion pair recombination
yields.
\end{abstract}
\maketitle
\section{Introduction}
Radical-ion pairs have been at the focus of chemical,
biological and physical research for the last few decades, mostly
because of their relevance to photosynthesis \cite{boxer_photo}.
It is a chain of electron-transfer reactions \cite{boxer_rev_1}
that converts the photon energy absorbed by pigment molecules in
the photosynthetic reaction center to energy capable of driving
further chemical reactions of biological importance. The
radical-ion pairs formed in this intricate photochemical machinery
further complicate the picture, since the spin interactions of
their unpaired electrons make the whole process
magnetic-sensitive. The importance of the magnetic sensitivity of
photosynthetic reactions \cite{hab1,schulten_weller,boxer_rev_2}
was realized early on and intensely studied theoretically as well
as experimentally. The timing of photosynthetic reactions
apparently rests on the complicated interplay between
non-radiative electron-transfer rates, spin-state mixing rates and
spin-selective recombination rates, to name a few of the
participating processes. It is therefore not surprising that
theories and experimental techniques from all of the
aforementioned disciplines have been summoned to address this
problem.

Parallel to these developments, it was early on realized
 \cite{schulten1,schulten2} that the magnetic sensitivity of radical-ion pair (RIP)
recombination reactions might be at the basis of the biochemical
magnetic compass used by several species for navigation in earth's
magnetic field \cite{w1}. After the development of extensive theoretical
models \cite{hore1,ritz1,weaver,solovyov} and several supporting experiments \cite{mouritsen,ritz2,maeda},
the so-called RIP magnetoreception mechanism is now established as forming the
basis of the biochemical compass, at least of migratory avian
species.

The magnetic sensitivity of RIP reactions \cite{brock} can be
easily understood: the RIP is formed from a photoexcited
donor-acceptor dyad molecule (DA) in the singlet spin state.
Magnetic interactions with the external magnetic field and
hyperfine interactions with the molecule's nuclear spins induce
the so-called singlet-triplet (S-T) mixing. Since the
recombination of the RIP is spin-selective, i.e. the singlet
$^{\rm S}$(D$^{+}$A$^{-}$) RIP can recombine to the neutral DA
molecule, whereas the triplet $^{\rm T}$(D$^{+}$A$^{-}$)
recombines, if at all, to a triplet metastable excited state
$^{\rm T}$DA, it is seen that the magnetic field can directly
influence the end-result of the photoexcitation-recombination
cycle.

What has gone unnoticed so far, however, is the fact that the
spin-selective charge recombination process is fundamentally a
quantum measurement \cite{kom_prl}. The tunneling of the RIP into
a vibrational excited state of the DA molecule is a scattering
process the outcome of which depends on the RIP'a spin state. In
other words, the intermolecular recombination dynamics
continuously interrogate the radical-ion pair's spin state. This
interrogation can be readily described by quantum measurement
theory \cite{braginsky} and leads to purely quantum phenomena such
as the quantum Zeno effect \cite{misra}. The theoretical
description of magnetic-sensitive RIP recombination processes was
so far based on phenomenological classical reaction theory
\cite{steiner} which masked the quantum effects that are dominant
in the regime of low-magnetic fields. This is the reason that
classical theories were rather successful, i.e. most measurements
were performed at high magnetic fields, where classical reaction
theories are valid. Several puzzling experimental observations at
low fields were however left unexplained. As will be shown in the
following, the quantum Zeno effect crucially affects the relevant
time constants of the RIP recombination process, and naturally
leads to an understanding of several experimental findings that
could not be reconciled with classical RIP reaction theories. In
this article we will present the detailed quantum theory of the
RIP recombination process and elucidate the fundamentally
different interpretation of the low-magnetic-field effects that
emerges. We will outline the connection of the full quantum theory
to be presented with the classical reaction theories. We will also
provide an explanation of experimental findings that have
challenged the classical theory, in particular measurements
regarding the effect of deuteration on the magnetic-sensitivity of
RIP recombination yields.
\section{The quantum description of RIP recombination}
In Figure 1 we diagrammatically describe the evolution of the RIP
recombination. The RIP is at time $t=0$ created in the singlet
state $^{\rm S}$(D$^{+}$A$^{-}$), which coherently mixes with the
triplet $^{\rm T}$(D$^{+}$A$^{-}$) under the unitary action of the
magnetic hamiltonian ${\cal H}$. At the same time, the charge
recombination is performing a quantum measurement of the singlet-
and triplet-state projection operators $Q_{S}$ and $Q_{T}$, with
the measurement rate being $k_{S}$ and $k_{T}$, respectively. At
some point the measurement is over. If the measurement result is
$q_{S}=1$ (obviously meaning that $q_{T}=0$), the RIP can tunnel
to an excited vibrational state of the neutral DA molecule, $^{\rm
S}$(DA)$^{*}$, and from there decay to the DA ground state at a
rate $\gamma_{S}$. In the event that the measurement produces the
result $q_{S}=0$ (obviously meaning that $q_{T}=1$), the RIP
similarly ends up at a metastable triplet state of the DA
molecule, $^{\rm T}$DA, and this happens at a rate $\gamma_{T}$.
In this work we will not deal with the decay rates $\gamma_{S}$
and $\gamma_{T}$, since they not are relevant for the magnetic
dynamics.
\begin{figure}
 \centering
 \includegraphics[width=8 cm]{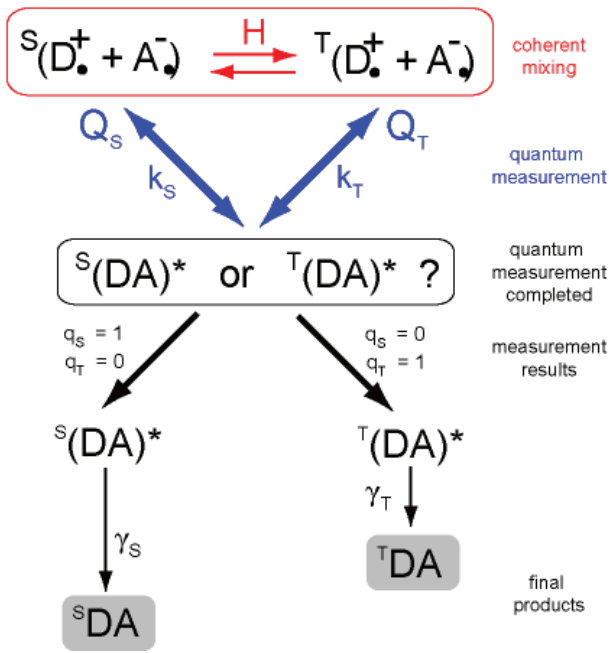}
 \caption{Quantum measurement scheme of the radical-ion pair charge recombination process. The coherent evolution of the RIP induced by
 the magnetic Hamiltonian is interrupted by the measurements performed by the charge-recombination process. At some random time the measurement
 is completed and a definite measurement result is available, so that the recombination can proceed accordingly.}
 \label{scheme}
\end{figure}

Quantum measurement theory \cite{braginsky} describes the effect
of a continuous measurement of a system observable $q$ performed
at a rate $k$ on the system's time evolution by the Liouville
equation: \beq \frac{d\rho}{dt}=-i[{\cal H},\rho]-k[q,[q,\rho]]
\eeq where $\rho$ is the quantum system's density matrix and the
first term describes the unitary evolution due to a Hamiltonian
interaction ${\cal H}$. In the case of the RIP recombination
process, the measured system observables are $Q_{S}$ and $Q_{T}$.
Therefore, the RIP's density matrix evolution equation will be
\beq \frac{d\rho}{dt}=-i[{\cal
H},\rho]-k_{S}[Q_{S},[Q_{S},\rho]]-k_{T}[Q_{T},[Q_{T},\rho]]\label{ev1}
\eeq Measurements performed on a coherently evolving quantum
system always cause decoherence. That is, \eqref{ev1} describes
the evolution of an initially pure state of the RIP to a mixed
state, during which the coherence induced by the Hamiltonian
${\cal H}$ is also dissipated. This is due to the
measurement-induced back-action. The RIP magnetic Hamiltonian to
be considered is comprised of the Zeeman interaction of the two
unpaired electrons with the external magnetic field and by the
hyperfine interactions with the molecule's nuclear magnetic
moments. We will consider for simplicity only two nuclei, with
nuclear spins $I_{1}$ and $I_{2}$, and different hyperfine
couplings (in frequency units) $A_{1}$ and $A_{2}$ to each of the
electrons. We will consider isotropic hyperfine interactions, as
we are not interested in describing directional effects
\cite{hore3} in this work. The magnetic Hamiltonian thus reads
\beq {\cal H}=h_{1}+h_{2} \eeq where \beq h_{j}=\omega
s_{jz}+a_{j}\mathbf{I}_{j}\cdot\mathbf{s}_{j}\label{ham12} \eeq is
the interaction Hamiltonian for each electron and $j=1,2$. The
Larmor frequency is \beq \omega=g_{s}\mu_{B}B=2.8 {\rm {{MHz}\over
{G}}}B[{\rm G}] \eeq and throughout we are using units such that
$\hbar=1$. For earth's field of $\sim$ 0.5 G, $\omega=0.7~{\rm
{\mu s^{-1}}}$. In the following, $\omega$ will be the unit of
frequency, i.e. we will normalize all rates by $\omega$. The
normalization is performed by $\omega$ and not by the hyperfine
frequency scale $a_{1}$ (or $a_2$) for the following reason. If
the mechanism is to serve for the measurement of the magnetic
field $\omega$, with precision $\delta\omega$, then from
Heisenberg's time-energy uncertainty relation follows that in a
measurement time $\tau$, the achievable precision $\delta\omega$
is such that $\tau\delta\omega\sim 1$. The precision
$\delta\omega$ is understood to be the width of the measured
distribution of frequencies $\omega$. Since the measurement time
$\tau$ is determined by the decay rate $\lambda$ of the RIP S-T
mixing, $\tau\sim 1/\lambda$, the relative width of the measured
distribution is $\delta\omega/\omega\sim \lambda/\omega$.
Therefore Heisenberg's uncertainty poses the requirement that
$\lambda/\omega<1$, if the mechanism is to be sensitive to the
magnetic field itself. If however, the signal-to-noise of the
actual detection process is $s$, than the measurement sensitivity
is enhanced to $\lambda/s\omega$. For example, if the
magnetic-field change is detected by the change in the recombined
molecules yield and the latter can be measured at the level of
1\%, then $\delta\omega/\omega\sim 0.01$ if $\lambda/\omega\sim
1$. It is obvious that the condition $\lambda/\omega<1$ can be
more easily satisfied at high fields, since the classical
recombination rates $k$ (and hence the decay rates $\lambda$) are
usually much larger than 1 ${\rm \mu s^{-1}}$. In the following,
we are going to show how the quantum Zeno effect allows the
existence of long S-T lifetimes, so that the condition
$\lambda/\omega\sim 1$ is satisfied even if $k\gg 1~{\rm \mu
s^{-1}}$.

We will now derive some basic properties of \eqref{ev1}. We denote
by $S=\tr\{\rho Q_{S}\}$ and $T=\tr\{\rho Q_{T}\}$ the probability
to find the RIP in the singlet and triplet spin state,
respectively. Due to the fact that $Q_{S}$ and $Q_{T}$ are
orthogonal projection operators, we have $Q_{S}^{2}=Q_{S}$,
$Q_{T}^{2}=Q_{T}$, $Q_{S}Q_{T}=0$ and $Q_{S}+Q_{T}=1$. Therefore,
using the identity $\tr\{[A,B]\}=0$, it follows from \eqref{ev1}
that
\begin{align}
\frac{dS}{d\tau}=-i\tr\{[Q_{S},{\cal H}]\rho\}\\
\frac{dT}{d\tau}=-i\tr\{[Q_{T},{\cal H}]\rho\}
\end{align}
and \beq \tr\{\rho\}=S+T=1 \eeq Does the fact that $\tr\{\rho\}=1$
mean that the RIP never recombines? Not at all. Every time the
RIP's spin state approaches the point $S=1$ (or $T=1$), there is a
non-zero probability of singlet- or triplet channel charge
recombination. If the RIP recombines, it ceases to exist in the
Hilbert space defined by the eigenstates of ${\cal H}$, and
together with is ceases the meaning of the density matrix $\rho$
and its normalization. This is formally described by a quantum
stochastic process \cite{gisin, wiseman} taking place in an
enlarged Hilbert space. A brief analysis has appeared in
\cite{kom_prl} and a more detailed presentation of the RIP's spin
state time evolution will be presented in a forthcoming
manuscript. Such a description requires the complete solution of
the density matric equation \eqref{ev1}. In the following we will
present some general arguments that do not necessitate such a
solution.
\section{Eigenvalue spectrum of the Liouville equation}
Since the nuclear spin multiplicity of the problem we are
considering is $(2I_{1}+1)(2I_{2}+1)$, and the electron spin
multiplicity is 4, the density matrix is a square $N$-dimensional
matrix, where $N=4(2I_{1}+1)(2I_{2}+1)$. Equation \eqref{ev1} thus
represents a set of $N^2$ coupled first-order differential
equations. We can rewrite them in the form $dR/dt=AR$, where
$R=(\rho_{11}~\rho_{12}~...~\rho_{NN})^{T}$ is a column vector of
dimension $N^2$ containing all density matrix elements and $A$ is
a $N^2$-dimensional square matrix. The eigenvalues of $A$ are
playing a central role in the considerations of this work, since
they are governing the time evolution of the RIP's spin state.
These eigenvalues are found by numerically diagonalizing the
matrix $A$, and in general are complex numbers of the form
$-\lambda+i\Omega$, where $\lambda\geq 0$ is termed the decay rate
and $\Omega$ the mixing frequency of this particular eigen-mode.
There are $n=N^2$ such eigenmodes. For example, if $I_{1}=1/2$ and
$I_{2}=1$, then $n=576$. It is easily seen that $n$ grows
exponentially with the number of participating spins, and the
numerical diagonalization quickly becomes a formidable problem.
This is the reason we limit the calculations to the case of just
two nuclear spins.
\begin{figure}
 \centering
 \includegraphics[width=8 cm]{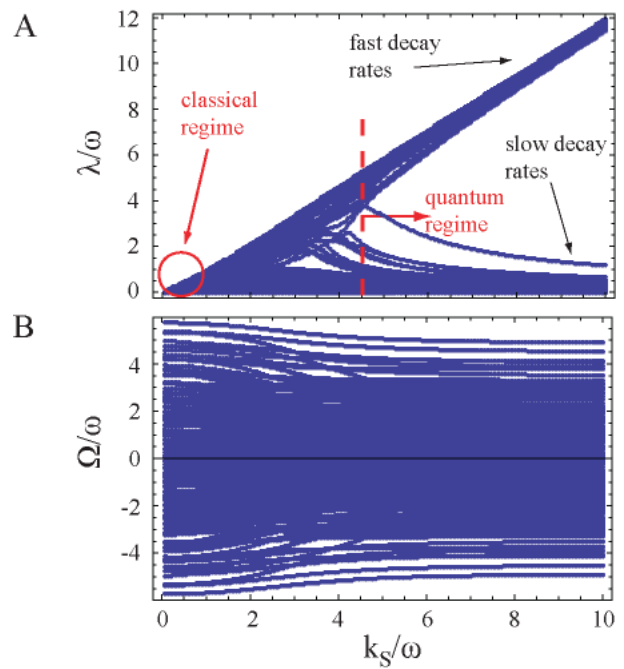}
 \caption{The decay rates and mixing frequencies of \eqref{ev1} are shown as a function of $k_{S}/\omega$, for $k_{T}=0.2k_{S}$,
 $I_{1}=I_{2}=1/2$, and $a_{2}=2a_{1}=3\omega$.}
 \label{fig:eig}
\end{figure}

A result of such a diagonalization is shown in Figure 2 for
typical values of the problem's parameters. We assume that
$k_{T}<k_{S}$, so that the recombination rate scale is determined
by $k_{S}$. We will distinguish two physically distinct regimes:
(i) $k_{S}/\omega\ll 1$, and (ii) $k_{S}/\omega\gg 1$. We term the
first regime classical, since it is in this case that classical
reaction theories are valid, whereas the second is termed quantum,
because it is in this parameter regime that the quantum Zeno
effect is at play. Before examining the magnetic sensitivity in
these two regimes, we note the following. Since the time evolution
of the density matrix is described by $n$ eigenmodes, every
density matrix element can be written as \beq
\rho_{ij}(t)=\sum_{l=1}^{n}{c_{ij}^{(l)}e^{(-\lambda_{l}+i\Omega_{l})t}},
\eeq where the coefficients $c_{ij}^{(l)}$ are determined from the
initial conditions. Therefore \beq \langle
Q_{S}(t)\rangle=\sum_{ij}{\rho_{ij}(t)q_{ji}}=\sum_{l=1}^{n}{e^{(-\lambda_{l}+i\Omega_{l})t}}A_{l}\label{QSt}
\eeq where $\rho_{ij}=\langle i|\rho|j\rangle$ and $|i\rangle$
with $i=1,...,n$ are the basis states, and \beq
A_{l}=\sum_{ij}{c_{ij}^{(l)}q_{ji}} \eeq with $q_{ji}=\langle
j|Q_{S}|i\rangle$. The reality of $\langle Q_{S}(t)\rangle$ is
based on the appearance of positive and negative mixing
frequencies, as shown in Figure 2B.
\subsection{Classical Regime}
We here study the magnetic sensitivity of RIP reactions when
$k_{S}/\omega<1$, i.e. in the regime of small recombination rates
and/or large magnetic fields. This is termed the classical regime.
We first note that in this regime the decay rates $\lambda$ scale
proportionally to $k_{S}$, $\lambda\sim k_{S}$. Secondly, the
mixing frequencies are constant, i.e. independent of the
recombination rate $k_{S}$. Hence all $n$ terms contribute to
$\langle Q_{S}(t)\rangle$, i.e. $n_{\rm slow}=n$ in \eqref{QStsl}.
In this case, the magnetic sensitivity of $\langle
Q_{S}(t)\rangle$ stems from the dependence of $q_{ji}$ on
$\omega$, i.e. a change in $\omega$ will result in a change of
$A_{l}$. As can be seen in Figure 3B, the matrix elements $q_{ij}$
and hence $A_{l}$ are varying strongly for fields $\omega/a>1$
until the very high field regime $\omega/a\gg 1$, where the
magnetic sensitivity drops to zero. In this high-field regime, the
${\rm T}_{\pm}$ branches of the energy eigenvalues have separated
away and there is small mixing only between ${\rm T}_{0}$ and S,
as can be seen in Figure 3A.
\begin{figure}
 \centering
 \includegraphics[width=8 cm]{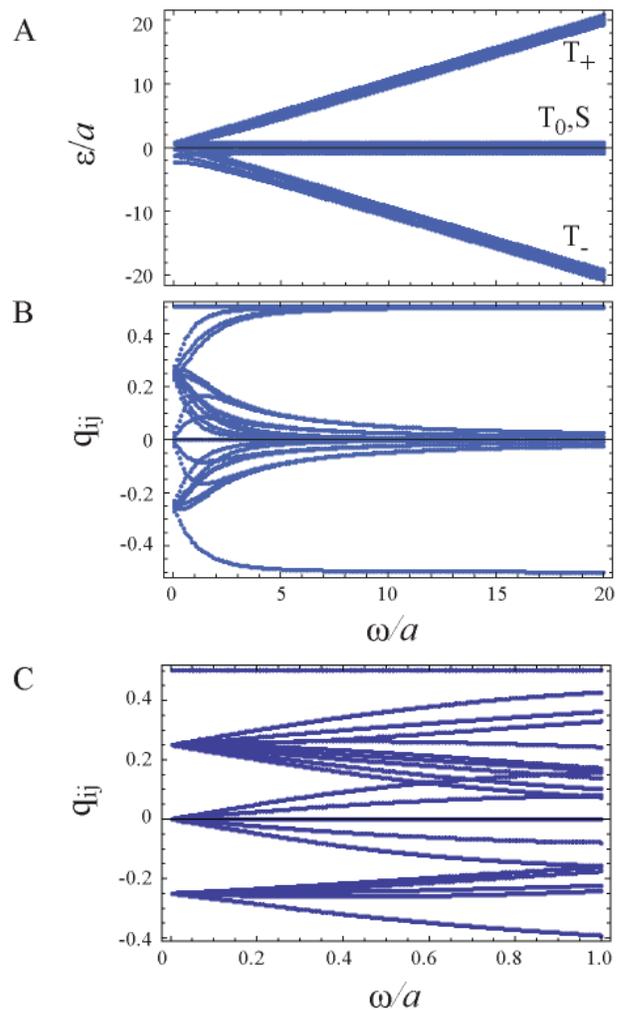}
 \caption{(A) Energy eigenvalues of the magnetic Hamiltonian for nuclear spins $I_{1}=I_{2}=1/2$ and hyperfine couplings
 $a_{1}=a$ and $a_{2}=2a$.(B) The corresponding matrix elements $q_{ij}$.
 In this case the $Q_{S}$-matrix is 16-dimensional, so there are
 256 matrix elements $q_{ij}$ of $Q_{S}$ in the basis formed by
 the eigenvectors of ${\cal H}$. It has been checked that the
 eigenvalues of the $Q_{S}$-matrix in this representation are 0
 and 1. (C) Zoom in at low magnetic fields, in order to better display the change of $q_{ij}$ in this regime.}
 \label{energies}
\end{figure}
The above considerations form the classical interpretation of
magnetic field effects on RIP recombination reactions, that have
already been analyzed in detail \cite{hore1,ritz1} (and references
therein).
\subsection{Quantum Regime}
The situation changes dramatically as the ratio $k_{S}/\omega$
increases. Indeed, it is seen from Figure 3C that for low magnetic
fields, $\omega/a\ll 1$, a 10\% change in $\omega$ produces a mere
2\% change in $q_{ij}$. According to classical theory, this is
further suppressed by the ratio $\omega/k_{S}$, resulting in a
negligible change of $\langle Q_{S}(t)\rangle$. However,
significant recombination yield changes are indeed observed
experimentally.
\begin{figure}
 \centering
 \includegraphics[width=8 cm]{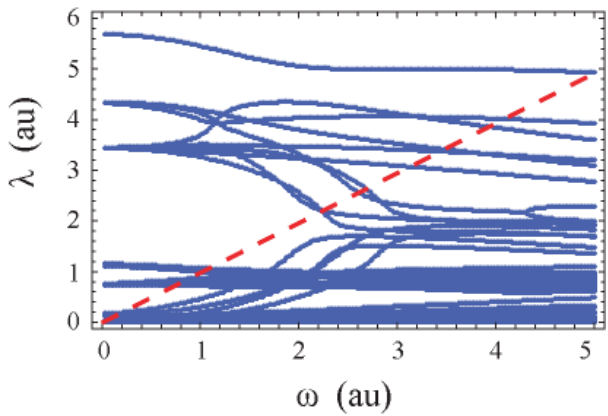}
 \caption{(A) Decay rates only the lower branch is shown) as a function of $\omega$ for $k_{S}=15$, $k_{T}=0.2k_{S}$, nuclear spins
 $I_{1}=I_{2}=1/2$ and hyperfine couplings
 $a_{1}=1.5$ and $a_{2}=3$. The dashed red line is the $\lambda=\omega$ line. The magnetic sensitivity derives from the changing number
 of decay rates crossing this line.}
 \label{eig2}
\end{figure}
The reason is that in this regime quantum effects become dominant,
fundamentally altering the nature of the magnetic sensitivity of
the recombination process. As seen in Figure 2A, it is in this
regime that the decay rates split into two branches, the fast
(upper branch) and slow (lower branch) decay rates. The latter are
a manifestation of the quantum Zeno effect \cite{misra}, the
physical significance of has been already described elsewhere
\cite{kom_bj}. We will here elaborate on the new interpretation of
magnetic sensitivity of RIP reactions in this quantum regime. Let
us suppose that some of the decay rates $\lambda_{l}$ are slow,
i.e. $\lambda_{l}\ll\Omega_{l}$, and some other are fast, i.e.
$\lambda_{l}\gg\Omega_{l}$. Let us further suppose that there are
$n_{\rm slow}$ of the former and hence $n_{\rm fast}=n-n_{\rm
slow}$ of the latter. Thus $n_{\rm fast}$ terms in \eqref{QSt}
will quickly decay away, leaving the $n_{\rm slow}$ terms to
determine the long-time behavior of $\langle Q_{S}(t)\rangle$:
\beq \langle Q_{S}(t)\rangle_{t\gg1/\lambda_{\rm
fast}}=\sum_{l=1}^{n_{\rm
slow}}{e^{(-\lambda_{l}+i\Omega_{l})t}}A_{l}\label{QStsl} \eeq It
thus follows that in this regime it is mainly the change of the
number $n_{\rm slow}$ with the magnetic field $\omega$ that is
responsible for the magnetic sensitivity of $\langle
Q_{S}(t)\rangle$. This number is a strongly changing function of
$\omega$ as shown in Figure 4.
\section{Phenomenological Liouville Equations}
The phenomenological density matrix evolution equations used so
far are \beq \frac{d\rho}{dt}=-i[{\cal
H},\rho]-k_{S}(Q_{S}\rho+\rho Q_{S})-k_{T}(Q_{T}\rho+\rho
Q_{T})\label{classL} \eeq The decay rates of the eigenvalue
spectrum of \eqref{classL} are shown in Figure 5. It is seen that
all the decay rates scale in proportion to $k_{S}$, i.e. the
quantum Zeno effect has disappeared. This is expected, because
equation \eqref{classL} by design forces the normalization of the
density matrix, $\tr\{\rho\}$, to an exponential decay: \beq
\frac{d\tr\{\rho\}}{dt}=-k_{S}\langle Q_{S}\rangle-k_{T}\langle
Q_{T}\rangle\label{trace}\eeq There is an analogous situation in
atomic physics experiments: transit time broadening \cite{budker}.
When an atomic coherence (for example a ground state spin
coherence) of an atomic vapor is probed by a laser, the atoms stay
in the interaction volume (defined by the laser beam) for a
limited time $\tau_{tr}$, determined by the atomic velocity
distribution. The measured spin coherence lifetime will thus be
limited by $\tau_{tr}$, which will add to the spin-resonance width
by $1/\tau_{tr}$. This is the transit-time broadening, which would
be described by an equation similar to \eqref{trace}: \beq
\frac{d\tr\{\rho\}}{dt}=-\frac{\tr\{\rho\}}{\tau_{tr}}\eeq
\begin{figure}
 \centering
 \includegraphics[width=8 cm]{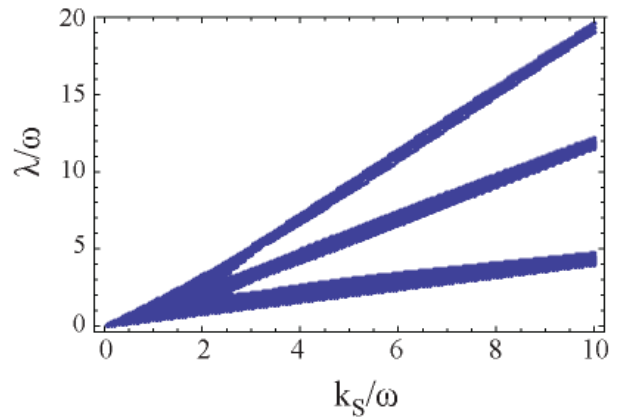}
 \caption{Decay rates of classical reaction theory Liouville equation \eqref{classL}. The calculation was done for $k_{T}=0.2k_{S}$,
 $I_{1}=I_{2}=1/2$, $a_{1}=1.5\omega$ and $a_{2}=3\omega$.}
 \label{classical}
\end{figure}
This equation phenomenologically describes the disappearance of
atoms with a time constant $\tau_{\rm tr}$. The situation with
radical-ion pairs, however, is fundamentally different, since in
this case there is a selective quantum measurement continuously
going on. An attempt to artificially introduce the crucial term
$Q_{S}\rho Q_{S}$ existing in \eqref{ev1} and missing from
\eqref{classL} was made in \cite{hab2} based on phenomenological
considerations, without any further consequences on the
theoretical description of RIP recombination dynamics. All
theoretical treatments, summarized in \cite{steiner}, have
thereafter used the phenomenological density matrix equation
\eqref{classL}.
\section{Explanation of Experimental Observations Inconsistent with Classical Reaction Theory}
We will now apply the interpretation of the low-field magnetic
sensitivity of RIP recombination reaction previously put forward
to address experimental observations that seem paradoxical within
the framework of classical reaction theory. Unusually long time
constants that have been observed in RIP reactions several times
have been explained elsewhere \cite{kom_prl}. Recently observed
\cite{maeda} magnetic sensitivity of RIP reaction rates which is
in stark contrast with classical reaction theory has been shown
\cite{kom_nat} to be fully consistent with the quantum theory.
Several experimental observations relevant to avian
magnetoreception based on magnetic-sensitive RIP reactions have
also been accounted for \cite{kom_bj}. Here we focus on some other
puzzling observations, namely the effect of the RIP's deuteration
on the magnetic sensitivity of the RIP recombination.
\subsection{Magnetic Sensitivity and Deuteration}
The effects of deuteration on the magnetic sensitivity of RIP
recombination reactions have early on \cite{blank} signalled the
inadequacies of classical reaction theories, as has been noted in
\cite{boxer_photo}. As stated in \cite{blank}, "The identical
triplet quantum yields in H and D samples does not support the
generally accepted idea that hyperfine interactions are
responsible for spin rephasing in the P870$^{+}$I$^{-}$ radical
pair". Indeed, it is rather straightforward to derive, based on
the classical spin-state mixing description, the result of a
change of hyperfine couplings. The triplet recombination yield is
crudely expected \cite{schulten_wolynes} to scale as $a/k_{S}$,
where $a$ is the dominant hyperfine coupling and $k_{S}$ the
singlet recombination rate. Any change in $a$ should thus lead to
a proportional change in the yield, contrary to the 1979
observations.

It is rather straightforward to explain this apparent
contradiction by use of the eigenvalue spectrum of the quantum
evolution equation \eqref{ev1}. Indeed, if we calculate the
eigenvalue spectrum for a molecule with $I_{1}=I_{2}=1/2$ and
hyperfine couplings $a_{1}=1.5\omega$ and $a_{2}=2.8a_{1}$, we
find that in the quantum regime ($k_{S}/\omega\gg 1$) the number
of decay rates $\lambda$ such that $\lambda/\omega<1$ are 58.2\%
of the total (256). If we repeat the calculation for $I_{1}=1/2$,
$I_{2}=1$, $a_{1}=1.5\omega$ and $a_{2}=0.86a_{1}$, i.e. we
replaced the second nucleus, which before was assumed to be a
hydrogen, with a deuterium, we find that the previous percentage
changes to 62.5\% (the total number of eigenvalues is now 576)
i.e. the relative yield change is only 7\%. However, the change in
the hyperfine coupling is about 300\%, clearly exemplifying the
disparity of classical expectations with observed data.
\begin{figure}
 \centering
 \includegraphics[width=8 cm]{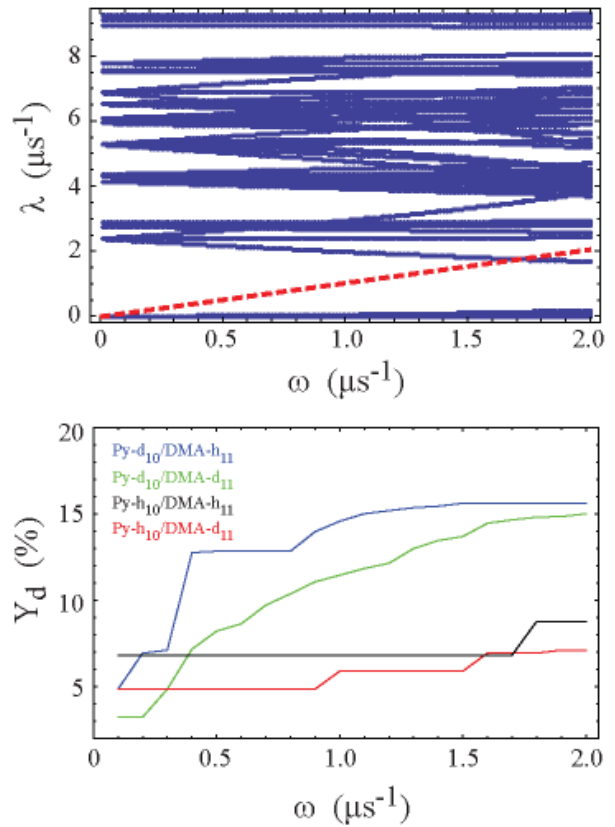}
 \caption{Magnetic sensitivity of the reaction yield of the Py-DMA radical-ion pair. What is calculated is the
 drop in the singlet-recombination yield. The magnetic Hamiltonian contains two nuclear spins, with $I_{1}$ and $I_2$ being either 1/2 or 1, dependent
 on which of the four RIPs is considered. The hyperfine couplings are taken from \cite{timmel_sc}. The singlet and triplet recombination rates used
 were $k_{s}=50~{\rm \mu s}^{-1}$ and $k_{T}=0.2k_{S}$, respectively. Similar results where obtained for $k_{S}=10,20~{\rm \mu s}^{-1}$.
 (A) The decay rates of Py-${\rm h}_{10}^{-}$/DMA-${\rm h}_{11}^{+}$ as a function of magnetic field. Similar plots are obtained for the other
 radical-ion pairs. The magnetic sensitivity is found by counting the decay rates that are smaller than $\omega$.
 (B) Estimated singlet-yield drop from the count of the slow decay rates for all four Py-DMA RIPs. It is seen that the protonated pyrene
 pairs show no magnetic response.}
 \label{pydma}
\end{figure}
We will now focus on more recent data regarding the effect
deuteration has on the magnetic sensitivity \cite{timmel_sc,hore2}
of the the pyrene (Py)-dimethylaniline (DMA) RIP recombination
dynamics. What has been observed is that the low-field-effect is
most pronounced when only one of the radicals exhibits strong
hyperfine couplings, with the second radical being deuterated and
hence having substantially reduced hyperfine couplings (the
magnetic moment of deuterium is about one third of the proton's).
In particular, in the case of protonated pyrene RIPs Py-${\rm
h}_{10}^{-}$/DMA-${\rm h}_{11}^{+}$ and Py-${\rm
h}_{10}^{-}$/DMA-${\rm d}_{11}^{+}$, there was almost no
magnetic-sensitivity of the exciplex fluorescence. On the
contrary, the recombination dynamics of deuterated pyrene RIPs
Py-${\rm d}_{10}^{-}$/DMA-${\rm h}_{11}^{+}$ and Py-${\rm
d}_{10}^{-}$/DMA-${\rm d}_{11}^{+}$ showed a clear low-field
sensitivity. In Figure 6 we show the calculated behavior of the
magnetic sensitivity. For each of the four RIPs we calculate the
eigenvalue spectrum and count the decay rates $\lambda$ that are
smaller than $\omega$. This spectrum is shown for one of the RIPs
in Figure 6A. Half the percentage of those decay rates is a good
estimate \cite{kom_bj} for the decrease in the singlet
recombination yield. It is seen that the relative yield change with
magnetic field displayed in Figure 6B is in perfect agreement with
observations. The absolute values of the yield drop
are not accurately predicted due to the simplicity of the two-spin
model used (they also depend on the particular value of $k_{S}$).

In summary, we have analyzed the physical interpretation of low
magnetic field effects that is bestowed upon the radical-ion pair
recombination reactions by their full quantum-mechanical
description. The explanatory power of this theory has been applied
in understanding experimental observations on the magnetic
sensitivity of RIP reactions that were conflicting classical
reaction theories. It has been shown that the eigenvalue spectrum
of the density matrix equation describing the time evolution of
the RIP's spin state accounts for the observations regarding RIP
deuteration. In particular, the slow decay rates imposed by the
quantum Zeno effect on the eigenvalue spectrum, and their
dependence on the problem's physical parameters, are shown to
contain rich dynamical information. Finally, it is rather crucial
to address the effect of low magnetic fields on photosynthetic
reactions under this new perspective.

\end{document}